# Sliding friction and superlubricity of colloidal AFM probes coated by tribo-induced graphitic transfer layers

Renato Buzio*,[†], Andrea Gerbi[†], Cristina Bernini[†], Luca Repetto[‡], Andrea Vanossi[§,⊥]

[†] CNR-SPIN, C.so F.M. Perrone 24, 16152 Genova, Italy

[‡] Dipartimento di Fisica, Università degli Studi di Genova, Via Dodecaneso 33, 16146 Genova, Italy

[§] CNR-IOM Consiglio Nazionale delle Ricerche - Istituto Officina dei Materiali, c/o SISSA, Via Bonomea 265, 34136 Trieste, Italy

[⊥] International School for Advanced Studies (SISSA), Via Bonomea 265, 34136 Trieste, Italy

**ABSTRACT:** Colloidal probe Atomic Force Microscopy (AFM) allows to explore sliding friction phenomena in graphite contacts of nominal lateral size up to hundreds of nanometers. It is known that contact formation involves tribo-induced material transfer of graphite flakes from the graphitic substrate to the colloidal probe. In this context, sliding states with nearly-vanishing friction, i.e. superlubricity, may set in. A comprehensive investigation of the transfer layer properties is mandatory to ascertain the origin of superlubricity. Here we explore the friction response of micrometric beads, of different size and pristine surface roughness, sliding on graphite under ambient conditions. We show that such tribosystems undergo a robust transition towards a low-adhesion, low-friction state dominated by mechanical interactions at one dominant tribo-induced nanocontact. Friction force spectroscopy reveals that the nanocontact can be superlubric or dissipative, in fact undergoing a load-driven transition from dissipative stick-slip to continuous superlubric sliding. This behavior is excellently described by the thermally-activated, single-asperity Prandtl-Tomlinson model. Our results indicate that upon formation of the transfer layer, friction depends on the energy landscape experienced by the topographically-highest tribo-induced nanoasperity. Consistently we find larger dissipation when the tribo-induced nanoasperity is sled against surfaces with higher atomic corrugation than graphite, like $MoS_2$ and $WS_2$, in prototypical Van der Waals layered hetero-junctions.

**INTRODUCTION**

Superlubricity loosely refers to situations in which sliding friction vanishes or very nearly vanishes.[1] For graphitic systems, such as graphite and graphene flakes,[2–6] characterized by weak dispersive interactions, superlubricity develops because of the presence of atomically flat shear planes at the contacting interfaces. In typical experimental setups the superlubric regime emerges after the built-in of tribo-induced material, called transfer layer (TL), from the graphite substrate onto the sliding countersurface.[2,7–9] Nanosized flakes forming the graphitic TL,[10] similarly to other 2D materials (e.g. $MoS_2$),[11] are thought to randomly arrange in misaligned configurations, providing an effective cancellation of lateral force components via interface incommensurability (structural lubricity), which might lead to friction coefficients as low as $10^{-3}$. In this context, one groundbreaking experiment, dating back to 2004 by Dienwiebel *et al.*,[2] demonstrated interlayer registry-dependent friction in a nanoscale contact between a graphite flake attached to an AFM tip and a pristine graphite substrate. Specifically, it was shown that when the two contacting atomic lattices are aligned in registry high interlayer friction values are measured, whereas friction practically vanishes for orientationally misaligned configurations. Structural lubricity was later documented in other graphitic nanosystems (e.g. nanoribbons and nanotubes)[12] and is nowadays actively investigated in Van der Waals layered hetero-junctions, where the lattice mismatch between graphite and other crystalline counter-surfaces might help to realize stable incommensurate sliding.[13–16]

Recently, colloidal AFM experiments have addressed the scale up of structural lubricity from nanoscale to mesoscale. It has been shown that when ideally smooth silica microspheres (nominal curvature radius $R{\sim}10\mu m$, surface roughness $\sigma < 2$nm) are repeatedly rub on Highly Oriented Pyrolitic Graphite (HOPG) under ambient conditions, the contact may evolve towards a low-adhesion, low-friction state due to the formation of a graphitic TL on the beads' surface.[10,17] As for the case of sharp AFM probes,[2,18,19] concurrent evidences from scanning probe and high-resolution electron microscopies indicate that few-layer and multilayer graphene flakes, generated by fracture and peeling of graphite at the surface-exposed step edges, do form the TL on the colloidal probes. In particular the tribo-transferred flakes mostly retain the crystalline phase[10] and lubricity[17] of bulk graphite, and ultra-small friction coefficients ($\lesssim 10^{-3}$) have been measured. To ascertain the microscopic mechanisms governing superlubricity in such case, the contact mechanics at the TL/HOPG interface needs to be elucidated. Early reports[10] speculated that lubricity relies on the pristine surface roughness of the colloidal probe, that incorporates nanoasperities covered by randomly oriented graphene nanopatches. This implies a multi-asperity contact mechanics, exploiting the cumulative effect of many incommensurate nanocontacts. However, large-scale simulations for the contact of rough spheres reveal that an ubiquitous single-asperity regime virtually emerges at any contact interface under low-loads conditions.[20] This is because at the lowest loads the interfacial mechanics of a rigid rough microparticle (pressed against an ideally-smooth surface) is governed by the first (nano)asperity that touches. Recently, we found that single-asperity contact mechanics – involving one dominant graphitic nanocontact – provides a comprehensive explanation of the friction behavior and superlubricity displayed by smooth silica microspheres ($R\sim12\mu m$, $\sigma<1$nm) sliding on graphite.[17] There is certainly need to better clarify through experiments how common microparticles' properties, such as their size and roughness, affect the formation of the superlubric TL.

In this paper we go beyond the case of an ideally smooth silica microparticle,[10,17,21] by considering a representative set of nominally-rigid colloidal probes having different curvature radius ($R\sim 2-25\mu m$) and pristine surface roughness ($\sigma\sim 0.5-15$nm). We complement conventional AFM force spectroscopy with knowledge derived from the TL morphology and atomic-scale friction force spectroscopy. We show that the phenomenology underpinning the appearance of superlubricity is robust against variations of the microparticle size and pristine surface roughness. Indeed, it is the roughness associated to the tribo-induced TL that makes the TL/HOPG contact nanoscopic in size, so that superlubricity may appear depending on the specific energy landscape experienced by the topographically-highest tribo-induced nanoasperity. Thanks also to numerical modeling, we attest the excellent agreement of atomic-scale friction data with the thermally-activated, single-asperity Prandtl-Tomlinson model. Noteworthy, these findings appear highly relevant to ascertain the origin of interfacial friction in prototypical Van der Waals hetero-junctions between the tribo-induced graphitic TLs and other bulk layered materials. In this respect the single-asperity Prandtl-Tomlinson model offers clear pathways to predict dissipation of hetero-contacts with typical transition metal dichalcogenides, like $MoS_2$, $WS_2$ and $NbSe_2$ single-crystal substrates. Also, results can provide a better comprehension of the mechanisms that assist the emergence of dry superlubricity by carbon-based additives.[5,22,23]

**EXPERIMENTAL SECTION**

**Colloidal probes.** We used two sorts of colloidal probes: 'small ones' with nominal diameter $\leq 11\mu m$, made of borosilicate glass (BG), and 'large ones' with diameter $> 20\mu m$ made of silica ($SiO_2$) or titanium (Ti). The small colloidal probes were commercially-available products (Novascan Technologies). According to the manufacturer, they consisted of rectangular shaped cantilevers with BG beads of diameter $\sim 5\mu m$ (cantilever elastic constant $k_C \sim 4.5$N/m) and $\sim 11\mu m$ ($k_C \sim 14$N/m) respectively. We prepared the larger colloidal probes using a micromanipulation stage (Newport M-460A-XYZ 3 axis stage equipped with SM/DM-13 differential screws) coupled with an optical microscope[24]. We glued monodisperse microspheres of $SiO_2$ (MicroParticles GmbH) or Ti (Alfa Aesar GmbH)[25] to rectangular Si cantilevers (MikroMasch HQ:NSC35/tipless, NSC12/tipless) using molten Shell Epikote resin dispensed on heated probes (heating temperature $\sim 140°C$, 50W heater cartridge controlled by a Lake Shore 335 temperature controller). The elastic constant of each cantilever was measured by Sader's method[26] before the bead attachment, and was in the range $k_C \sim 2.5 - 5.8$N/m. The surface morphology of the beads was routinely characterized in the course of experiments by reverse AFM imaging on a spiked grating (Tipsnano TGT01). The surface roughness $\sigma$ was evaluated from topography maps, and after subtraction of the spherical shape, as the standard deviation of the distribution of heights:

$$\sigma = \sqrt{\langle (h - \langle h \rangle)^2 \rangle} \quad (1)$$

where $h$ is the local surface height and $\langle \ \rangle$ represents the statistical average over the AFM-measured $h$ values. Additionally, SEM imaging of the beads' surface was completed under a 1kV acceleration voltage (CrossBeam 1540 XB by Zeiss). Figure 1 shows the pristine surface morphology of the four representative probes used in the present study. For the BG beads (Figure 1a,b) their cross sections are satisfyingly fitted by circular fits, with curvature radii ($R = 1.9\mu m$ and $4.1\mu m$) slightly smaller than those declared by the manufacturer. The surface morphology of the larger BG bead is indeed far from homogeneous, as several isolated bumps – from tens to hundreds of nanometers in size – are superimposed over the bead circular profile. Surface roughness for the BG probes, evaluated after subtraction of the spherical shape, is $\sigma \sim 4$nm ($R = 1.9\mu m$) and $\sigma \sim 15$nm ($R = 4.1\mu m$) on areas of $\sim 1\mu m^2$. On the other hand, the pristine $SiO_2$ colloidal probe ($R = 11.0\mu m$) is a nearly-ideal smooth sphere, with surface roughness $\sigma \sim 0.5$nm (Figure 1c). Finally, the surface of the Ti colloidal probe ($R = 25.0\mu m$) appears rather irregular, with nanometric scratches and bumps, and surface roughness $\sigma \sim 6$nm (Figure 1d).

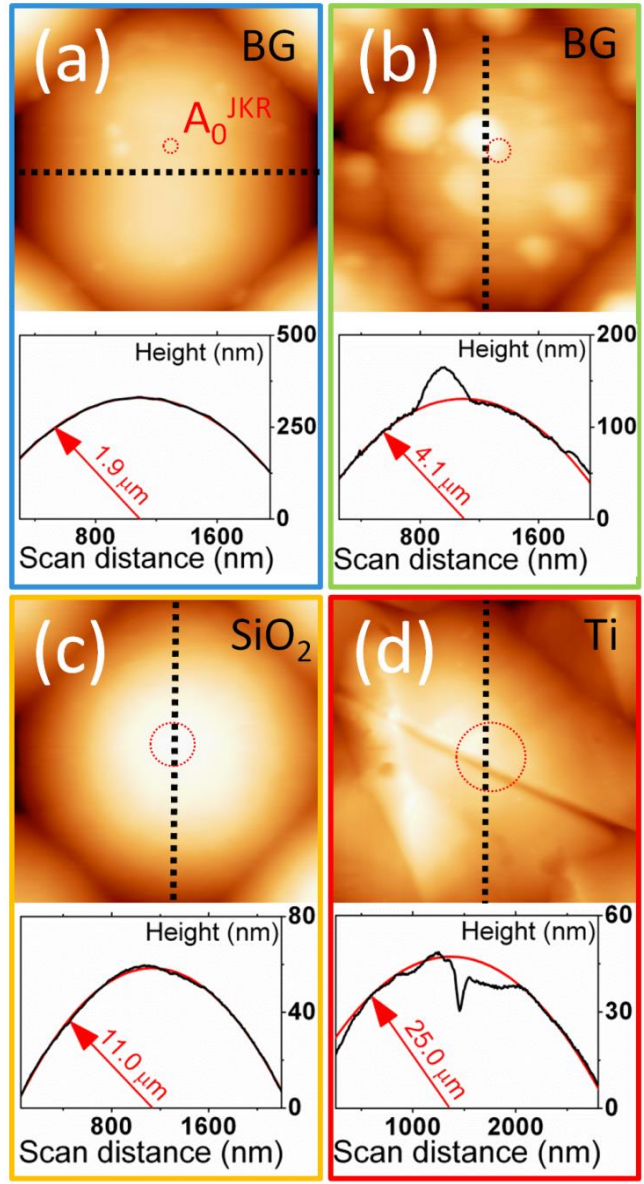

**Figure 1.** (a),(b) Reverse AFM topographies of two BG colloidal probes in their pristine state. Bottom panels show the cross-sectional height (black line), the circular fit (red line) and the estimated curvature radius $R$. (c) As in (a), for a SiO$_2$ probe. (d) As in (a), for a Ti probe. Red dotted circles in (a-d) show how the zero-load contact area $A_0^{JKR}$ scales with $R$ (see text).

Beads of different size and degree of surface roughness were intentionally chosen to evaluate their impact on TL formation and superlubricity. According to Johnsons-Kendall-Roberts JKR theory (see Figure S1 of Supporting Information), ideally-smooth spheres with the same radius $R$ of the colloidal probes span a zero-load Hertzian contact area $A_0^{JKR}$ ranging between $\sim 8 \times 10^3 \text{nm}^2$ and $\sim 2 \times 10^5 \text{nm}^2$ when placed in contact with graphite. This corresponds to a mesoscopic contact diameter $\sim 100 nm - 500$nm (see red dotted circles in Figure 1). We expect that both the BG/HOPG ($R = 1.9$μm) and the SiO$_2$/HOPG interfaces should have (initial) zero-load contact areas close to $A_0^{JKR}$, in view of the small pristine roughness. On the contrary $A_0^{JKR}$ only provides an upper limit to the real contact area at the (rougher) BG/HOPG ($R = 4.1\mu m$) and Ti/HOPG interfaces. Again, these differences offer the possibility to address the contact formation under different working conditions.

**Substrates.** For HOPG, we used a square substrate of grade ZYB (MikroMasch), with the armchair and zigzag crystallographic directions of the (0001) HOPG surface parallel to the edges of the square.[27] Graphite was cleaved by adhesive tape in ambient air prior to friction measurements. The substrate was mounted onto the sample stage with a fixed orientation that corresponded to a sliding direction slightly off ($\leq 10°$) the armchair crystallographic direction. Commercially-available crystals of 2H-MoS$_2$ and 2H-WS$_2$ (from HQ Graphene), together with a laboratory-grown 2H-NbSe$_2$ single-crystal sample (see Figure S2 for the typical surface topography of this sample),[28] were also used for friction measurements after being cleaved by adhesive tape in air. The cleaved surfaces usually showed atomically-flat terraces extending several micrometers.

**Friction force experiments.** Sliding friction experiments were performed under ambient conditions (relative humidity RH$\sim 40 - 60\%$, temperature T = $23 \pm 3°C$) by means of a commercial AFM operated in contact mode (Solver P47-PRO by NT-MDT, Russia). The normal force $F_N$ and the lateral force $F_L$ were calibrated as reported elsewhere.[17] Friction force *vs* normal load ($F_f$ *vs* $F_N$) characteristics were obtained by decreasing $F_N$ every ten lines from a large value ($F_N \sim 400 - 700$nN) to the pull-off point; the corresponding lateral forces (calculated from the difference between forward and backward scans) were averaged in between the normal force $F_N$ jump to produce one data point.[22,29] The $F_f$ *vs* $F_N$ curves were acquired over atomically-flat areas free of atomic steps (scan range from $5 \times 5$nm$^2$ up to $300 \times 300$nm$^2$). Friction maps were analyzed with National Instruments LabView and they were displayed using WSXM.[30] Normal force *vs* displacement curves were obtained by recording the cantilever deflection (i.e. the normal force $F_N$), while ramping the relative distance between the tip and sample, and they were transformed into normal force *vs* distance ($F_N$ $vs$ $D$) curves by assigning $D = 0$ to the region where normal force becomes strongly repulsive.[22,31] Adhesion force was estimated at the pull-off point.

**Analysis and simulations of atomic-scale friction data.** Experimental atomic-scale friction maps were analyzed in the framework of the one-dimensional Prandtl-Tomlinson (PT) single-asperity model.[32] In the PT modeling approach, a single point-like asperity (mimicking the AFM tip) is elastically driven over an on-site (sinusoidal) potential of amplitude $E_0$ and lattice spacing $a$ (the corrugated substrate) via a pulling spring, connecting the position of the probing tip itself to an external stage (the AFM support) moving at constant speed. The spring value $k$, in the PT description, represents an effective system stiffness, combining both the torsional features of the cantilever and the mechanical properties of the interface contact. The PT model describes two distinct dynamical regimes for the tip motion depending on the Tomlinson parameter $\eta$, representing the ratio between the corrugation and the effective elastic energy. When $\eta \leq 1$ the system total potential exhibits only one minimum and the sliding motion is smooth with vanishing energy dissipation; for $\eta > 1$, two or more potential minima appear in the energetic landscape and the dynamics becomes intermittent and characterized by dissipative stick-slip transitions. Similarly to [32,33], we evaluated $E_0$, $\eta$ and $k$ as a function of applied load. In short, at each normal load $F_N$, we selected lateral force traces with periodicity $a$ (~$0.21 - 0.25$nm for HOPG; ~$0.29 - 0.32$nm for MoS$_2$ and WS$_2$), corresponding to trajectories characterized by individual slip jumps approximately along the zigzag crystallographic direction. Frictional traces exhibiting different slips were disregarded. For each selected force profile, the corrugation amplitude $E_0$, the Tomlinson parameter $\eta$ and the effective contact stiffness $k$ were estimated as:

$$E_0 = \frac{aF_{L,max}}{\pi} \qquad (2)$$

$$\eta = \frac{2\pi F_{L,max}}{ak_{exp}} - 1 \qquad (3)$$

$$k = \frac{\eta+1}{\eta} k_{exp} \qquad (4)$$

where $F_{L,max}$, $a$ and $k_{exp}$ are, respectively, the highest sampled local force maxima, the slip distance, and the lateral force slope at the initial stick phase. Experimentally, for each applied load $F_N$, the statistically meaningful mean values of $E_0$, $k$ and $\eta$ were obtained by averaging over several lateral friction traces. Theoretically, friction *vs* displacement profiles to be compared with experiments, were extracted by integrating *via* a 4th-order Runge-Kutta algorithm the underdamped Langevin equation for the PT model:

$$m\ddot{x} + m\gamma\dot{x} = -\frac{\partial V(x,t)}{\partial x} + \xi(t) \qquad (5)$$

where the potential energy of the system $V(x,t)$ reads:

$$V(x,t) \equiv -\frac{E_0}{2}\cos\left(\frac{2\pi x}{a}\right) + \frac{1}{2}k(vt-x)^2 \qquad (6)$$

The instantaneous lateral friction trace was simply evaluated by considering the time-dependent elongation of the pulling spring:

$$F_L = k(vt - x) \qquad (7)$$

For the effective tip mass we used multiples of $m_0 = 1 \times 10^{-12}$Kg, whereas the Langevin damping $\gamma = 2\sqrt{k\,m}$ was chosen to reproduce experimental force traces. The thermal noise term $\xi(t)$ satisfies the fluctuation-dissipation theorem: $\langle \xi(t)\xi(t') \rangle = 2m\gamma k_B T \delta(t-t')$. Temperature T = 296K, spring stiffness $k$ and sliding velocity $v = 30$nm/s where chosen to match AFM experiments.

**RESULTS AND DISCUSSION**

**TL formation from nanoscale force spectroscopy.** According to previous studies,[10,17,21] the 'ultralow-friction' interface results from tribo-induced material transfer from the graphitic substrate to the sliding countersurface. We enabled material transfer by driving colloidal probes over HOPG with the sliding direction perpendicular to the local orientation of the surface steps (scan areas from $1 \times 1 \mu m^2$ to $8 \times 8 \mu m^2$, velocity $\sim 1 - 50 \mu m/s$ ). This choice in fact aids material transfer from HOPG to the AFM probe.[21] As resumed in Figure 2, the process of TL formation showed qualitatively similar features for all beads. Figure 2a reports the evolution of the friction force for each probe, displayed over an arbitrary sliding distance of $\sim 2.5 \times 10^4 \mu m$. One observes that initially, in the absence of the TL, friction force fluctuates randomly in the range of several hundreds of nN (high-friction state). As scanning progresses however, a permanent drop of the friction force to a few nN takes place (low-friction state). The fall of friction force is a typical signature for the formation of a graphitic TL. Below we use TL/HOPG to indicate the TL-mediated contact between each colloidal probe and HOPG. We note that the sliding distance to achieve the ultralow-friction TL/HOPG interface varied greatly from probe to probe.[10,17] This reflects the random character of interfacial wear and tribotransfer processes at the base of the TL formation.[17] It is known that nanoscale asperities superimposed on the beads surface can initiate fracture and peeling of graphite at the surface exposed step edges at lateral forces above $F_L \sim 50nN$.[34] Therefore, wear is likely to take place for the pristine beads ($F_L \gg 100nN$) and flakes can be tribotransferred from HOPG to the contact area. Indeed this was already documented by *ex-post* AFM micrographs of the HOPG abrasive wear, specifically showing the presence of worn graphite monolayers and layered debris (i.e. few-layers graphene flakes) for smooth $SiO_2$ colloidal probes sliding across step edges.[17] Figure 2b reveals that the TL formation was also accompanied by a simultaneous drop of contact adhesion. Specifically the large adhesion at the pristine interfaces, $F_A \sim 1.7 - 2.2 \mu N$ for $R \geq 11 \mu m$ and $F_A \sim 250 - 500nN$ for $R \leq 4.1 \mu m$, was reduced to $F_{A,TL} < 200nN$ for the TL/HOPG interface regardless of the probe radius $R$. The absence of a trivial scaling $F_{A,TL} \propto R$ after TL formation signals that a few random morphological features likely control contact mechanics[35] (see also next section). Furthermore, the abrupt snap-off-contact for the TL/HOPG interface supports the picture that contact forces are governed by one dominant contact.[36] Figure 2c shows a selected ensemble of $F_f$ vs $F_N$ characteristics acquired over atomically-flat areas of graphite (from $200 \times 200 nm^2$ to $500 \times 500 nm^2$) for the TL/HOPG contacts. Linear regression gives $\mu \sim 0.001$, which agrees with previous reports[37] and attests that an 'ultralow-friction' interface is achieved for each colloidal probe.

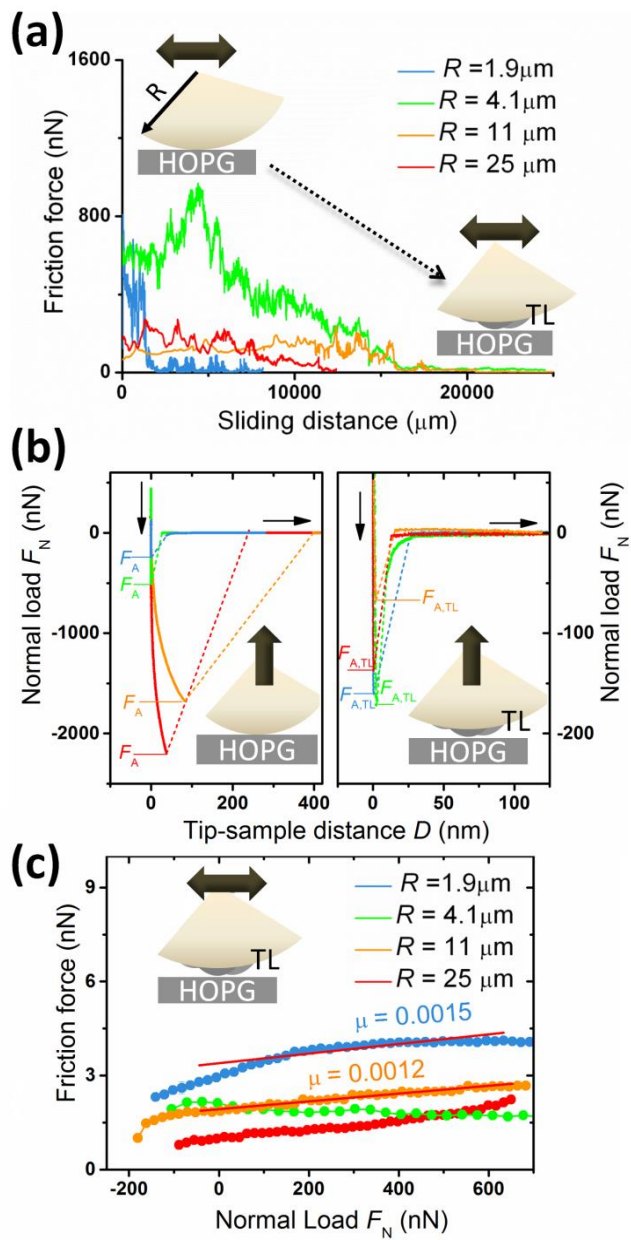

**Figure 2.** (a) Friction force *vs* sliding distance, showing the transition from a 'high-friction' state proper of the pristine colloidal probes to an 'ultralow-friction' state attesting the TL formation. (b) Normal load *vs* distance curves for the pristine

contacts (left) and for the TL/HOPG contacts (right; color code as in (a)). (c) Load-dependent friction characteristics for the ultralow-friction TL/HOPG contacts.

In the next section we show that the TL is rough on the nanometers scale, and in fact it leads to a substantial increase of the surface roughness $\sigma$ for each probe. Therefore, in line with our previous AFM experiments using ideally smooth colloidal probes,[17] we claim that the main effect of the TL is to turn the nominally mesoscopic contact area $A_0^{JKR}$ into a nanoscopic one. The roughness-induced decrease of the real contact area readily gives the breakdown of the friction force $F_f$, as the interfacial shear stress with graphite is known to not vary substantially with contact pressure.[22] Also, the capillary adhesion - that typically dominates over van der Waals forces under ambient conditions [38] - decreases if interfacial roughness overcomes the Kelvin length scale of capillary condensation $\lambda_K = 0.52$nm. Thus the TL roughness is expected to greatly suppress the contact adhesion too, i.e. $F_{A,TL} \ll F_A$.

**TL morphology and asperity-mediated contact.** We routinely characterized the evolution of the surface morphology of the colloidal beads during sliding friction measurements by means of reverse AFM imaging.[39] According to Figure 3, the TL consists of flakes of lateral size $\sim 50 - 100$nm, clustered within the contact region of each bead. Cross sections attest that the flakes rise tens of nanometers away from the beads' profile ($\sim 5 - 50$nm), in fact representing isolated asperities. Additionally, we confirmed *via* reverse friction maps the ultralow friction response of such flakes (not shown),[17] this being an indication of their graphitic nature. Hence, after TL formation, the beads surface is covered by superlubric nanoasperities. Such conclusions are well corroborated by high-resolution SEM micrographs of the graphitic flakes (see Figure S3).

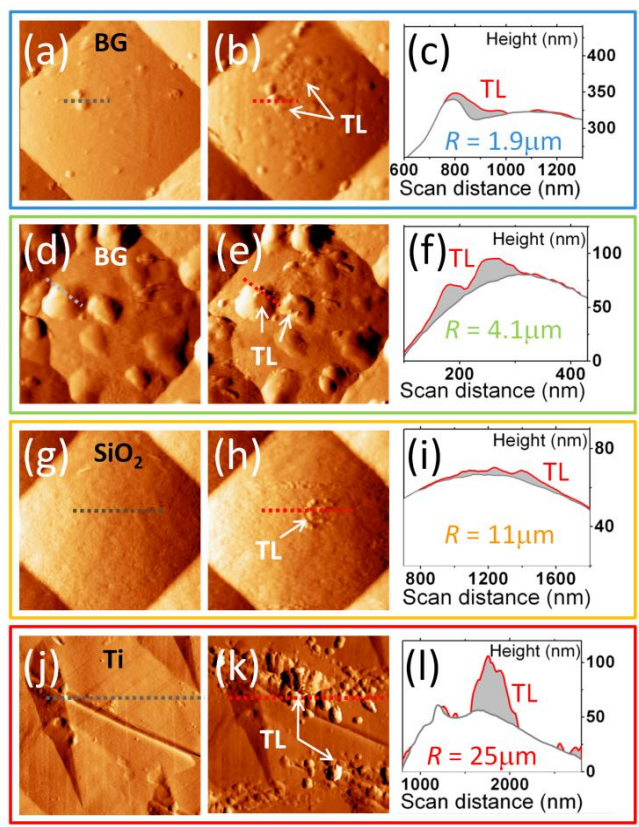

**Figure 3.** Representative AFM maps (error signals) comparing the surface of each bead respectively before ((a),(d),(g),(j)) and after ((b),(e),(h),(k)) the TL formation (curvature radius $R$ increases from top to bottom). Cross sectional heights ((c),(f),(i),(l)) - taken along the dashed lines in the AFM maps - shown that the TL originates several protruding nanoasperities, or it covers the already-existing ones.

Figure 4a documents the systematic increase of the beads' morphological roughness within the contact region after the TL formation. In particular, for the smooth $SiO_2$ bead ($R = 11$μm) roughness increases from $\sigma \sim 0.5$nm (without TL) to $\sigma_{TL} \sim 2$nm (with TL). Importantly, this value agrees with the thickness of individual nanoflakes imaged by AFM onto worn graphite areas.[17] Also, it agrees with the 2.3nm –thick TL measured by Li *et al.* by means of high-resolution electron microscopy, in colloidal probe AFM experiments using smooth silica beads.[10]

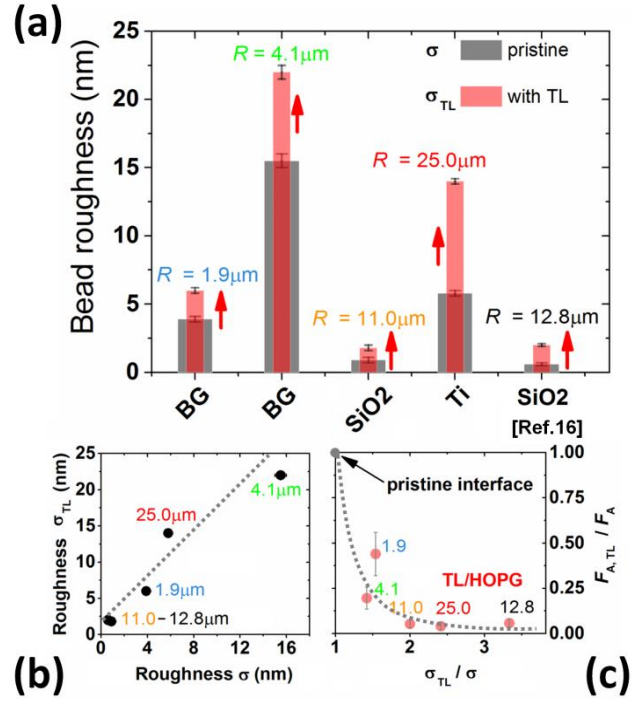

**Figure 4**. (a) Surface roughness measured within the contact region, respectively before ($\sigma$) and after ($\sigma_{TL}$) the TL formation: roughness increases for each colloidal bead (red arrows). Data from[17] are added for comparison. (b) Plot of $\sigma_{TL}$ vs $\sigma$, indicating good correlation of the two quantities (dash line is a guide to eye). (c) Plot of the normalized adhesion vs normalized roughness, showing the existence of a 'minimum adhesion plateau' for $\sigma_{TL} \sim 2\sigma$ ($\sigma_{TL}/\sigma = 1$ corresponds to pristine interfaces; dash line is a guide to eye).

The systematic increase of the roughness $\sigma_{TL}$ with the pristine roughness $\sigma$, resumed in Figure 4b, reflects the fact that the tribo-transferred flakes preferentially form new asperities rather than being incorporated within the already-existing beads' surface valleys, so that their overall effect is to enhance topographical variations within the contact region. This readily emerges by inspection of Figures 3d-e,j-k.

As mentioned in the previous section, the main role of the TL roughness is to decrease the real contact area between the beads and HOPG. As a result, both the pull-off (adhesive) force and the friction force are drastically reduced. The effects of surface roughness on adhesion have been studied extensively in particle technology, adhesion science, and fundamental physics and chemistry.[40] In the context of colloidal probe AFM experiments, the breakdown of adhesion upon increase of interfacial roughness is well documented for nominally rigid interfaces.[35,36,41,42] Adhesion reduction has been reported also when interfacial nanoasperities originate from accumulated wear debris,[43] in a situation similar to the tribo-transferred-flakes case. Noteworthy, a plot of the normalized adhesion ($F_{A,TL}/F_A$) vs normalized roughness ($\sigma_{TL}/\sigma$) (Figure 4c) indicates the existence of a 'minimum adhesion plateau' for the TL/HOPG interface (for $\sigma_{TL} \gtrsim 2\sigma$), in analogy with similar trends found for other systems.[35,36,42] This plateau corresponds to the adhesion experienced within a graphitic nanosized contact spot. By processing AFM topographies of the TL with threshold criteria, one gets a rough estimate of $\sim 0.5 \times 10^3 \text{nm}^3$ for the contact area at the topographically-highest contact spot (see Figure S4). Indeed, we have shown before using the Maugis-Dugdale contact mechanics theory,[17] that a contact area of about $\sim 10^3 \text{nm}^2$ corresponds to an interfacial adhesion $\gamma_{TL/HOPG} \sim 130 \text{mJ/m}^2$, which is not far from $\sim 160 - 220 \text{mJ/m}^2$ obtained from single-asperity AFM experiments on graphite. For a more detailed calculation of the adhesion force $F_{A,TL}$ including the contribution of capillary adhesion under 40-60% RH conditions, see elsewhere (section S4 in [17]).

We note that the occurrence of several superlubric asperities over the beads' surface does not necessarily imply a multi-asperity contact regime. To clarify this issue we observe that the TL formation was accompanied also by emergence of multiple-tip effects (Figure S5). These arise whenever the graphite roughness, associated to airborne adsorbates or to surface steps, has an aspect ratio comparable to the TL roughness.[44] Topographies, in such case, display repetitive patterns revealing the arrangement of those asperities affecting contact mechanics. On the contrary, normal and shear forces are likely supported by one asperity only (i.e. the most prominent one) when the probe slides over atomically-flat graphite.[17] This picture appears reasonable also with respect to scaling arguments derived from large scale molecular simulations.[20,42] Briefly, these indicate that the contact between a self-affine rough microsphere and a flat substrate evolves with the ratio $F_N/N_C$, where the critical load $N_C = (9\pi^3/16) E^* R^2 (h'_{rms}/\kappa)^3$ depends on material and geometrical parameters ($E^*$ is the contact modulus, $R$ is the sphere radius, $h'_{rms}$ is the dimensionless root mean square topographical slope and $1/\kappa$ is a dimensionless constant). According to the model of the nonadhesive hard-wall limit, for $F_N/N_C \gtrsim 1$ contact mechanics is single-asperity and dominated by the microscopic sphere radius $R$, for $10^{-4} \lesssim F_N/N_C < 1$ a multi-asperity regime takes place due to a statistical number of nanometric contact spots, whereas a crossover to a single-asperity regime (governed by the first nanoasperity that touches) occurs for $F_N/N_C < 10^{-4}$. In the present case one gets the estimate $N_C \sim 120 - 210 \mu N$ (with $R = 1.9\mu m$ and $11\mu m$, see Figure S6). This suggest that the TL/HOPG interface is nearby the cross-over from multi-asperity to single-asperity for $F_N \sim 100nN$, and it can easily evolve towards a single-asperity nanocontact in the small-loads limit (e.g. $F_N \sim 10nN$). Accordingly, neither the curvature radius $R$ nor the pristine surface roughness ($\sigma < \sigma_{TL}$) are expected to significantly impact the TL/HOPG contact mechanics.

**Atomic-scale friction and superlubricity of the tribo-induced TL sliding on graphite.** Analysis of atomic-scale lateral force maps allowed to characterize the elementary dissipation mechanisms at the TL/HOPG interface. Such maps typically showed stick-slip motion in the normal load range $100nN \lesssim F_N \lesssim 600nN$ (Figure 5a). Remarkably, this occurred for all colloidal probes. The ubiquitous presence of stick-slip motion signals the existence of an interlocking mechanism between one dominant TL nanoasperity and the underneath graphite subtrate.[17]

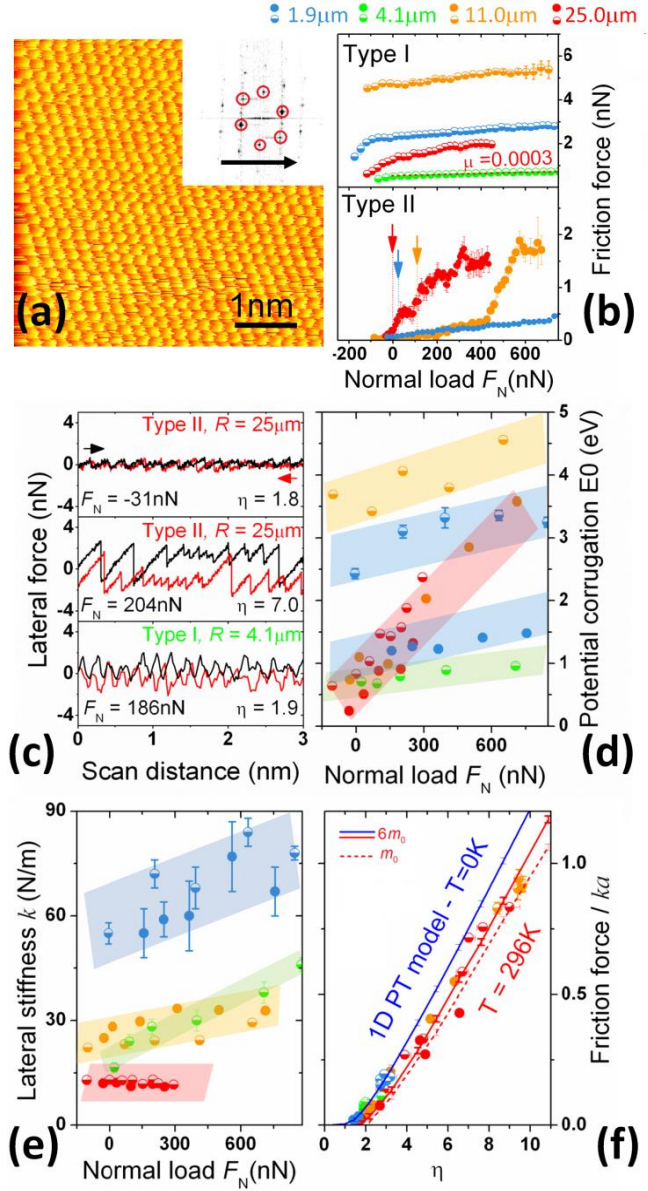

**Figure 5** (a) Friction map acquired at the atomic scale ($R = 25\mu m$); in the inset, 2D FFT with the hexagonal symmetry of the graphite lattice together with the probe sliding direction). (b) Representative set of $F_f$ vs $F_N$ characteristics ($v = 33nm/s$). Arrows indicate the critical load for the transition from superlubric sliding to stick-slip in Type II curves (see text). (d) Load-dependent friction loops for some of the friction curve in (b). (d-e) Load-dependent potential corrugation $E_0$ and contact stiffness $k$ estimated from analysis of curves in (b). (f) Normalized friction force $F_f^*$ vs $\eta$ and comparison with the 1D PT model ($m_0 = 1 \times 10^{-12}$Kg; see text).

The $F_f$ vs $F_N$ curves were affected by some variability at the atomic scale. This reflects changes in the fine details of the TL/HOPG contact, taking place during the experiments. Some representative characteristics are reported in Figure 5b, conveniently separated into two groups. On one hand there are curves with a weak dependence on load and finite friction force $F_f \geq 0.5$nN down the pull-off point ('type I' friction curves). On the other side, there are entirely different characteristics showing vanishing friction at the lowest loads and a monotonic increase at some point ('type II' friction curves). Notably, each colloidal probe showed characteristics of both types in the course of the experiments (with the exception of the BG probe with $R = 4.1$μm, that showed only 'type I' curves). The main features of such characteristics were not affected by the specific protocol we used to ramp $F_N$ (see Figure S7). Inspection of friction loops in 'type II' curves (Figure 5c) shows that energy dissipation substantially evolves with the normal load $F_N$, from nearly-frictionless continuous sliding (e.g. at $F_N = -31$nN) to highly-dissipative (single-slip and multi-slip) stick-slip motion ($F_N = 204$nN); this is not the case for type I curves, which did not show continuous superlubric sliding even at the largest tensile loads. Qualitatively this phenomenology agrees with atomic-friction experiments conducted on graphite by means of sharp AFM tips[45,46] and suggests to use the PT model to elucidate the friction behavior of the TL/HOPG interface for each colloidal probe. To this end, we estimated the contact parameters $E_0$, $k$ and $\eta$ of the PT model as a function of $F_N$. The analysis was carried out for each $F_f$ vs $F_N$ characteristic of Figure 5b. Results are summarized in Figure 5d-f. For each colloidal probe, the potential corrugation $E_0$ increases with $F_N$ (Figure 5d) and assumes the smallest values for 'type II' curves, whereas higher barriers are systematically associated to 'type I' characteristics. For each probe, 'type I' and 'type II' curves have comparable values of the lateral stiffness $k$ (Figure 5e). This analysis allows mapping different $F_f$ vs $F_N$ curves into the adimensional scaling $F_f^*$ vs $\eta$, where $F_f^* \equiv F_f/ak$ is the normalized friction force. In this way it turns out that all the experimental $F_f$ vs $F_N$ characteristics follow reasonably well the thermally-activated PT model (Figure 5f). Notably, the model predicts a friction force $F_f^*$ systematically smaller than for the athermal (T = 0K) case, as thermal energy do assists the contact asperity of the TL to jump from one minimum of the potential to the next.[33,47] We thus conclude that small (unpredictable) variations of the TL/HOPG contact move the tribosystem in different positions of the $F_f^*$ vs $\eta$ plot. In particular, 'type II' curves mainly populate the small dissipation region $1.7 \leq \eta \leq 6$. In this region, the contact transitions from continuous superlubric sliding to dissipative stick-slip. As $F_N$ increases the energy barrier $E_0$ increases faster than the contact stiffness $k$ (Figure 5d,e), so that the Tomlinson parameter $\eta \propto E_0/k$ grows with $F_N$, causing the transition from continuous sliding to a stick-slip mode.[46] On the other hand, curves of 'type I' populate mostly the region $\eta > 6$, where the TL/HOPG contact dissipates mechanical energy only by stick-slip instabilities.[46] One can thus appreciate that the faster growth of $E_0$ with $F_N$ shown by some of the contact junctions (e.g. the red stripe in Figure 5d) merely implies a larger modulation of the Tomlinson parameter η upon changes of $F_N$, but it does not capture the crucial difference between the two types of curves.

In the previous study[17] on ideally smooth silica beads, we already pointed-out the evocative correspondence between the phenomenology depicted in Figure 5 and friction duality, i.e. the existence of two distinct regimes of finite and vanishing friction, observed respectively in single-asperity AFM experiments[48] and nanoparticles manipulation experiments[49] on HOPG. In such reports vanishing friction was attributed to lattice mismatch within the contact area, while finite friction was ascribed to some pinning source, like interfacial defects[50] or airborne contamination. In line with such reasoning, the 'type II' behavior in Figure 5b likely reflects structural lubricity (i.e. in-plane misalignment) for a tribo-induced flake with respect to graphite, holding up to a critical normal force value where dissipative stick-slip sets in. Such threshold can be tensile ($F_N < 0$) or compressive ($F_N > 0$). In the last case it agrees with the $\sim 1 - 40$nN found for nanosized contacts.[32,45,51,52] In turn, the load-driven transition from superlubric sliding to stick-slip motion can reflect the progressive emergence of contact pinning effects. At this level of discussion, one can ascribe pinning to an increase of the degree of interfacial commensurability, e.g. due to small (reversible) in-plane rotations of the tribotransferred flake bearing contact,[33] or to load-induced pinning effects at the contact edges.[53] Such effects might explain both the monotonic rise of friction for the ultralow 'type II' $F_f$ $vs$ $F_N$ curve of Figure 5b above some critical load value. Following discussion above, 'type I' regime could rely on a single-asperity contact fully dominated by pinning effects. Our study does not strictly address the physical origin of pinning. Spectro-microscopy materials characterization might in principle shed light on this issue as well as drive tailored atomistic simulations, albeit the small size and sparse coverage of the tribo-trasferred flakes at the contact interface make a similar approach challenging. As the TL flakes are known to have an interlayer spacing of 0.38nm larger than the 0.33nm of pristine graphite,[10] oxygen-containing functional groups might intercalate flakes during substrate tribo-exfoliation. Therefore, we speculate that sliding-induced defects or airborne contaminants might contribute to pinning effects.

The single-asperity PT phenomenology at atomic scale (Figure 5) indicates that the friction response of the nominally mesoscopic TL/HOPG junction indeed depends on the energy landscape experienced by the topographically-highest tribo-induced nanoasperity. Given the $\sim 50 - 100$nm lateral size of the tribo-transferred flakes, such an asperity may involve one or at most a few neighbor flakes. Random interactions of the bead with the HOPG atomic roughness can drive reorganizations of such tribo-induced nanoasperity, that are at the base of the friction fluctuations and variability of experimental friction *vs* load curves documented above. We point out that the friction signals were stable whenever the colloidal beads were sled over atomically-flat graphite regions, which in fact allowed to routinely obtain atomically-resolved friction maps (as that shown in Figure 5a). On the contrary, lateral movement of the colloidal beads across atomic roughness (e.g. the atomic-scale surface-exposed step-edges) easily triggered TL reorganizations accompanied by abrupt jumps in the base level of the friction force. These transitions explain the emergence of different friction characteristics (of 'type I' and 'type II') for the very same probe. More details on the appearance of such friction force jumps - both in 2D lateral force maps and 1D friction traces - can be found elsewhere.[17]

**Friction of the tribo-induced TL sliding on transition metal dichalcogenides.** One of the main advantages offered by graphene-coated AFM probes is to enable the investigation of interfacial forces and mechanical dissipation in Van der Waals layered hetero–junctions. In fact, the practical realization of crystalline heterogeneous contacts with 2D materials is crucial to systematically perform fundamental force spectroscopy studies;[54] also, it is often claimed to be a necessary condition to realize stable incommensurate sliding and – possibly – structural lubricity between mismatched lattices.[13–16] Quite unexpectedly, however, experiments with graphene-coated probes sled against common transition metal dichalcogenides (e.g. $ReS_2, TaS_2, MoS_2, WS_2$) attest an increase of the friction force compared to the graphene/HOPG homo-junction.[19,55] This issue still needs to be fully elucidated, as it demands to rationalize the complex interplay of intrinsic properties (e.g. contact size and elasticity, interfacial corrugation, degree of incommensurability) and extrinsic factors (e.g. interfacial roughness, environmental contaminants, wear). Hereafter we provide deeper insight by probing the frictional response of the graphitic TL sliding against a few transition metal dichalcogenides. We used the following methodology. First, the nearly-ideally smooth $SiO_2$ colloidal bead was sled over HOPG until a superlubric TL/HOPG contact formed. As discussed above, such contact was characterized by $\mu \lesssim 10^{-3}$ and by atomic-scale continuous sliding for $F_N$ up to a few tens of nN (as for 'type II' characteristics in Figure 5b). Then, the very same colloidal probe was retracted from HOPG and immediately engaged on the atomically-smooth surface of a freshly-cleaved bulk crystal, namely $MoS_2$ (TL/$MoS_2$), $WS_2$ (TL/$WS_2$) or $NbSe_2$ (TL/$NbSe_2$). Prior to engage, the probe was carefully located by optical microscopy over a surface region free from visible surface steps; scan area was kept in the sub-micrometric range ($< 500 \times 500 nm^2$) and prominent interactions of the graphitic TL (covering the $SiO_2$ bead) with surface steps were avoided. Results – summarized in Figure 6 – confirmed a general robust increase of the friction force at all loads, for each hetero-junction, compared to the superlubric starting condition. Noteworthy, both the single-asperity contact mechanics and the PT model offer pathways to explain the response of such tribosystems.

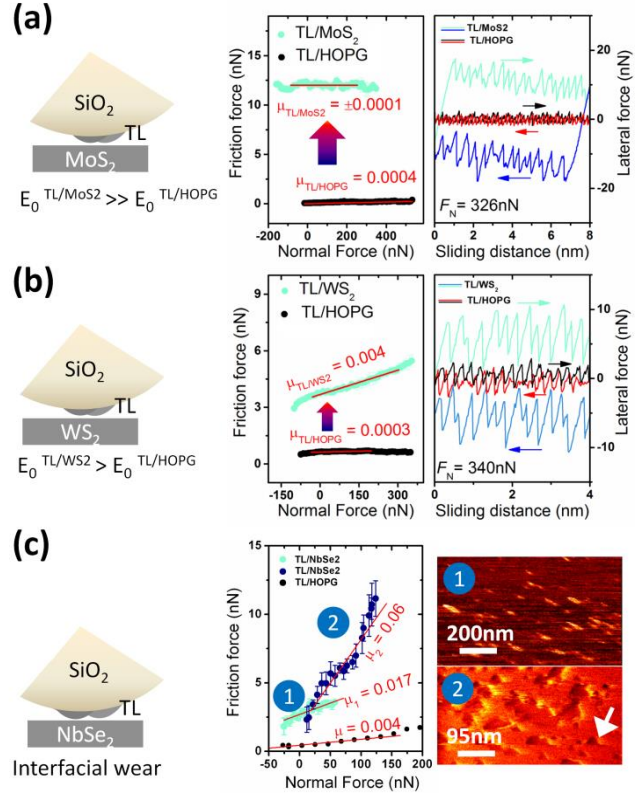

**Figure 6.** (a) Friction response of the TL/MoS$_2$ hetero-junction, resumed in terms of $F_f$ $vs$ $F_N$ curve (left panel) and atomic-scale lateral force traces (right panel). The higher friction of the hetero-junction compared to the TL/HOPG homo-junction reflects the substantial increase of the interfacial corrugation $E_0$. (b) As in (a), but for the TL/WS$_2$ hetero-junction. (c) Friction response of the TL/NbSe$_2$ contact, depicted in terms of $F_f$ $vs$ $F_N$ characteristics (left panel) and AFM maps (right panel). Interfacial wear dictates the friction increase (see text).

For the TL/MoS$_2$ hetero-junction, friction experiments were carried out up to a few hundreds of nN normal load (Figure 6a), without detection of surface wear. Typically, the friction force was larger ($F_f \sim 12nN$) than for the TL/HOPG superlubric case ($F_f \sim 0.1 - 0.5 nN$) and was almost load-independent down to the pull-off point ('type I' characteristic). Whereas this would result in a virtually zero friction coefficient ($\mu_{TL/MoS2} \sim \pm 10^{-4}$), atomic-scale friction data always indicated the occurrence of significant dissipative stick-slip with periodicity $\sim 0.3nm$ fitting the MoS$_2$ atomic lattice. This clearly excludes structural lubricity for our realization of such hetero-junction, involving far less extended tribocontacts if compared to other experimental realizations.[13] Indeed, according to the single-asperity PT model, dissipative stick-slip results from a Tomlinson parameter $\eta > 1$. This is reasonably expected for the TL/MoS$_2$ hetero-junction, since a nanoasperity sliding on MoS$_2$ is known to experience higher energy barriers than on HOPG [56] so that $\eta \propto E_0$ becomes substantially larger than 1. Consistently we found that the main effect of MoS$_2$ was to enhance interfacial corrugation from $E_0(\text{TL/HOPG}) \lesssim 0.8\text{eV}$ to $E_0(\text{TL/MoS}_2) \sim 8.0\text{eV}$ that ultimately led to $\eta(\text{TL/MoS}_2) \sim 8$ (see Figure S8). Likewise, for the TL/WS$_2$ hetero-junction, the friction force was higher ($F_f \sim 3-5\text{nN}$) than for the TL/HOPG counterpart ($F_f \sim 0.1-0.5\text{nN}$) and dissipative stick-slip with $\sim 0.3\text{nm}$ wavelength occurred (Figure 6b). Again, we found that $E_0$ increased from $E_0(\text{TL/HOPG}) \sim 1.0\text{eV}$ to $E_0(\text{WS}_2/\text{HOPG}) \sim 4.0\text{eV}$, which ultimately led to $\eta(\text{TL/WS}_2) \sim 6$. The condition $F_f(\text{TL/HOPG}) < F_f(\text{TL/MoS}_2)$ agrees with previous studies on nanosized layered hetero-junctions based on 2D flake-wrapped AFM tips,[19] whereas $F_f(\text{TL/WS}_2) < F_f(\text{TL/MoS}_2)$ can be ascribed to the different out-of-plane deformations of the MoS$_2$ and WS$_2$ surfaces.[57]

A different situation occurred for the TL/NbSe$_2$ hetero-junction (Figure 6c). Sliding friction experiments were initially carried out at moderate normal loads ($F_N \leq 50\text{nN}$) and showed an increase of both friction force and friction coefficient ($F_f \sim 1-5\text{nN}$; $\mu_1 = 0.017$) compared to the TL/HOPG contact ($F_f \sim 0.1-0.5\text{nN}$; $\mu_{\text{TL/HOPG}} \approx 0.004$). Remarkably, the lateral force often showed sharp spikes at randomly-located surface spots (top AFM map in Figure 6c). We realized *a posteriori* that such spikes were precursors of the initiation of atomic surface wear, with the AFM probe sweeping away material fragments from the at the NbSe$_2$ basal plane. In fact, in a second series of experiments conducted at higher loads ($F_N \leq 150\text{nN}$), there was an additional increase of the friction coefficient ($\mu_2 = 0.06 \gg \mu_{\text{TL/HOPG}}$) accompanied by the appearance of characteristic triangular-shaped holes in the investigated area (bottom AFM map in Figure 6c). The mean depth of the holes roughly corresponds to the thickness of one layer of NbSe$_2$. Wear phenomena on NbSe$_2$ have been observed previously[58] in AFM experiments using nanosized tips, with a normal load threshold that can be roughly located between $F_N \sim 25\text{nN}$ [58] and $F_N \sim 100\text{nN}$ [59,60] according to the actual tip size. Here, the observation of surface wear appears as an obvious consequence of the TL contact mechanics, because an individual tribo-induced nanoasperity mediates the contact between the colloidal bead and NbSe$_2$ and leads to a high contact pressure exceeding the limit of wearless sliding.

Experiments discussed above, far from being conclusive, strengthen the general correspondence between graphitic tribo-induced TLs and nanocontacts. Results for TL/MoS$_2$ and TL/WS$_2$ hetero-interfaces indicate that contact pinning prevails over interfacial incommensurability as soon as graphite is replaced with the other layered bulk substrates. Simulations and tailored experiments (e.g. probing the rotational anisotropy of friction in carefully prepared homo- and hetero-interfaces)[2,19,55] appear necessary at this stage to shed more light on the main physical ingredients that govern the evolution of the interfacial corrugation $E_0$.

**CONCLUSIONS**

In this study we addressed the manifestation of ultralow-friction sliding states, including superlubricity, for colloidal AFM probes sliding on graphite under ambient conditions. We have shown that neither the pristine surface roughness nor the curvature radius of the colloidal bead have a major impact on the phenomenology underpinning the appearance of superlubricity. Rather, graphitic flakes are tribo-transferred at the bead-graphite contact region where they form new asperities, so that contact mechanics and mechanical dissipation depend on the details of the topographically-highest, tribo-induced nanoasperity. This clarifies many single-asperity effects encountered in AFM imaging and force spectroscopy, including the dramatic reduction of adhesion and friction forces upon TL formation, the origin of multiple-tips convolutions effects, and the occurrence of atomic-scale load-controlled transitions from continuous superlubric sliding to stick-slip dissipative motion. Superlubricity strictly occurs only in specific cases, and in the low-load regime. The correspondence of atomic friction data with the well-established PT model, suggests that the friction force ultimately arises from the interplay between interfacial crystalline incommensurability and pinning effects (both at the contact edges or within the contact area). These findings also explain why interfacial friction increases when the tribotransferred flakes are placed in contact with transition metal dichalcogenides, such as MoS$_2$ and WS$_2$. Since nanocontacts experience higher sliding barriers on such materials than on graphite, pinning prevails over interfacial incommensurability and a higher dissipation takes place. Our work contributes to the ongoing research on structural lubricity, by offering an original framework of experimental and theoretical evidences suitable to analyse colloidal probe AFM experiments on Van der Waals layered junctions.

## ASSOCIATED CONTENT

The Supporting Information is available free of charge on the

ACS Publications web site at DOI:

Contact mechanics calculations with the Johnson-Kendall-Roberts JKR model (Figure S1), AFM/STM topographies of the laboratory-grown 2H-NbSe$_2$ sample (Figure S2), SEM micrographs of the contact area of each bead (Figure S3), contact area for the topographically-highest nanoasperity (Figure S4), multiple-tip effects originated by a rough transfer layer (Figure S5), evaluation of the average local slope and critical normal load from AFM topographies (Figure S6), effect of contact loading/unloading on TL/HOPG friction characteristics (Figure S7), evaluation of atomic-scale contact parameters for sliding hetero-junctions (Figure S8).

## AUTHORS INFORMATION

Corresponding Author

* E-mail: renato.buzio@spin.cnr.it (R.B.).

Notes

The authors declare no competing financial interest.

## ACKNOWLEDGEMENTS

This work was financially supported by the MIUR PRIN2017 project 20178PZCB5 "UTFROM - Understanding and tuning friction through nanostructure manipulation". A.V. acknowledges also support from ERC Advanced Grant ULTRADISS, contract No. 86344023.

# Sliding friction and superlubricity of colloidal AFM probes coated by tribo-induced graphitic transfer layers

Renato Buzio*,†, Andrea Gerbi†, Cristina Bernini†, Luca Repetto‡, Andrea Vanossi§,⊥

† CNR-SPIN, C.so F.M. Perrone 24, 16152 Genova, Italy
‡ Dipartimento di Fisica, Università degli Studi di Genova, Via Dodecaneso 33, 16146 Genova, Italy
§ CNR-IOM Consiglio Nazionale delle Ricerche - Istituto Officina dei Materiali, c/o SISSA, Via Bonomea 265, 34136 Trieste, Italy
⊥ International School for Advanced Studies (SISSA), Via Bonomea 265, 34136 Trieste, Italy

## Supplementary Information

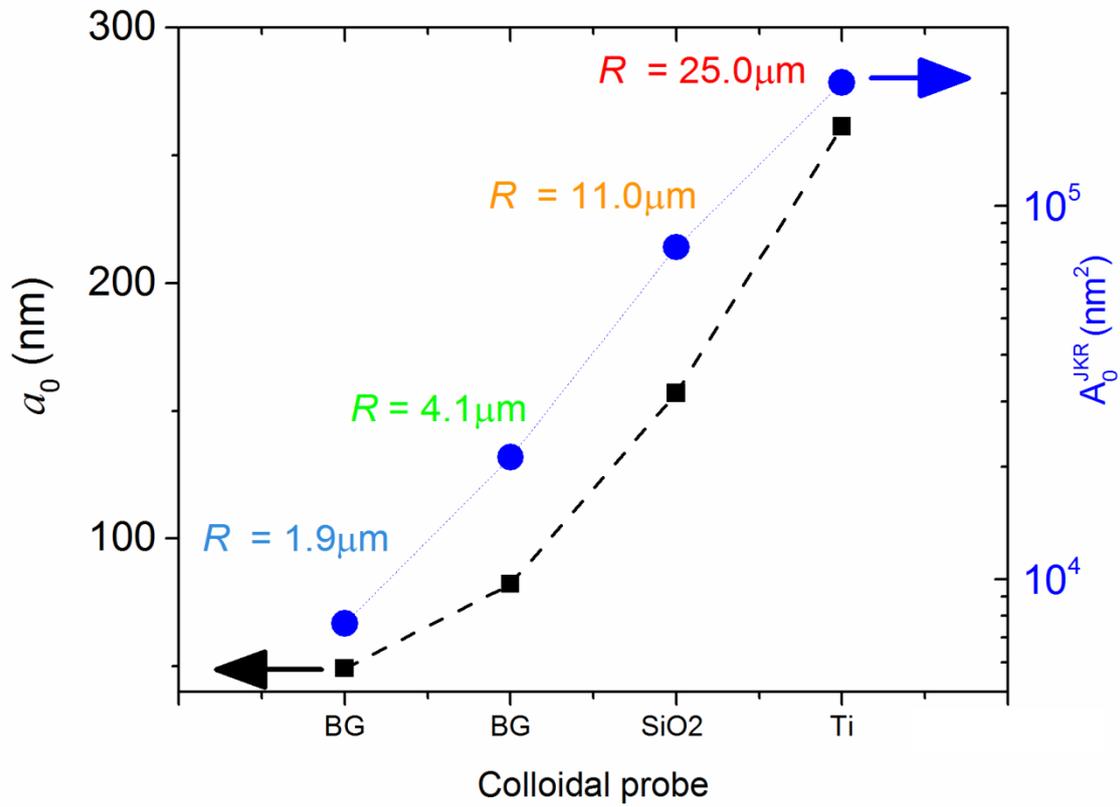

**Figure S1.** Zero-load contact radius $a_0$ and contact area $A_0^{JKR}$ evaluated using the JKR theory, for ideally smooth beads with the same size of those used in the present study (material parameters: Young modulus $E$ (BG) = 64GPa, Poisson ratio $\nu$(BG) = 0.2; $E$(SiO$_2$) = 70GPa and $\nu$(SiO$_2$) = 0.2; $E$(Ti) = 100GPa and $\nu$(Ti) = 0.35; $E$(HOPG) = 30GPa and $\nu$(HOPG) = 0.24; interfacial energy $\gamma$ = 50mJ/m$^2$). See also Ref. [17] of the main text for calculation details.

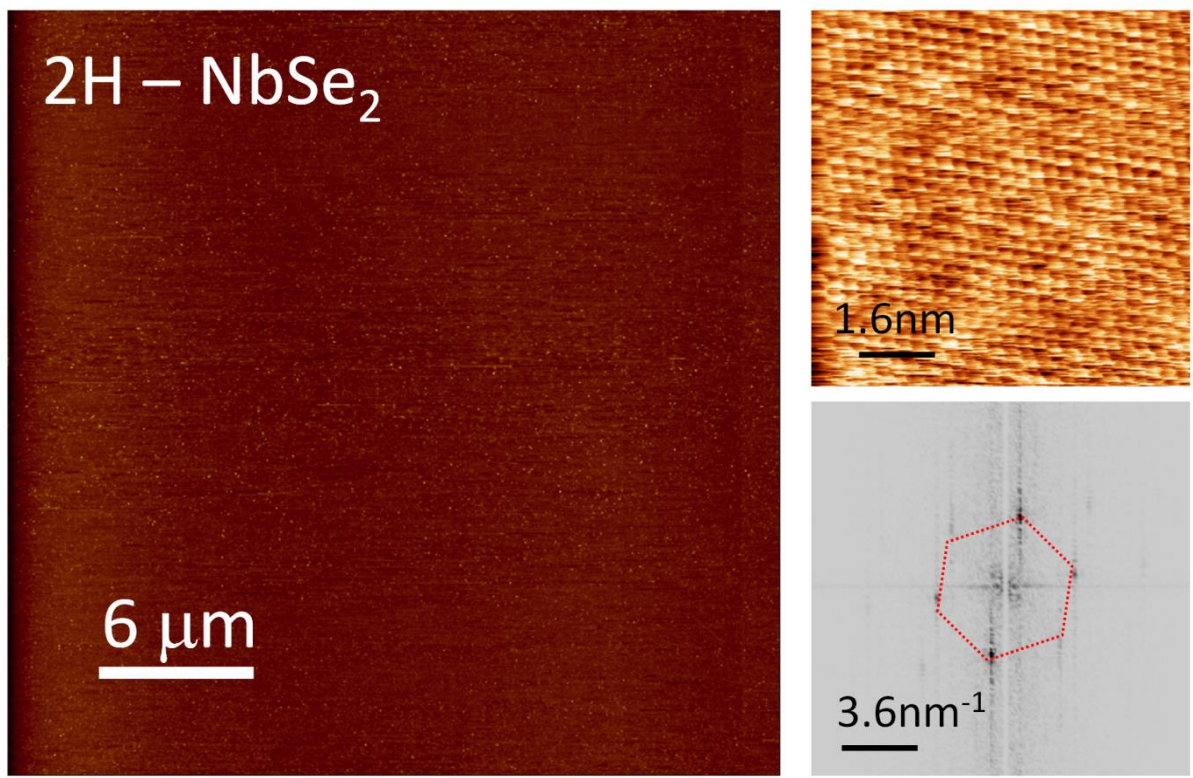

**Figure S2.** (Left) Typical AFM topography of the laboratory-grown 2H-NbSe$_2$ single-crystal sample: it shows an atomically-flat region, extending tens of micrometers in lateral size without atomic steps. (Right) Raw atomic-scale STM topography acquired on the same sample (Omicron Nanotechnology LT-STM, T = 80K). The hexagonal symmetry of the surface lattice is clearly resolved in the related 2D Fast Fourier Transform FFT map.

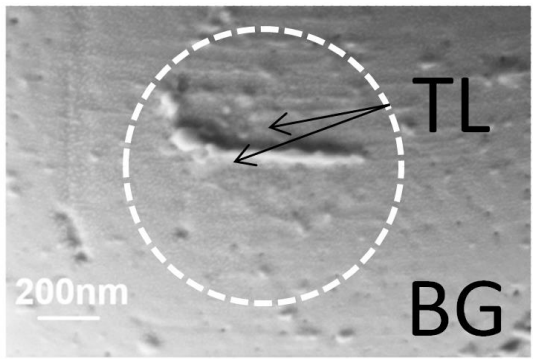
$R = 1.9\ \mu m$

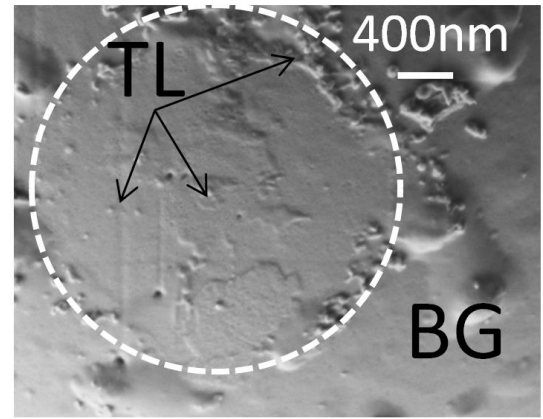
$R = 4.1\ \mu m$

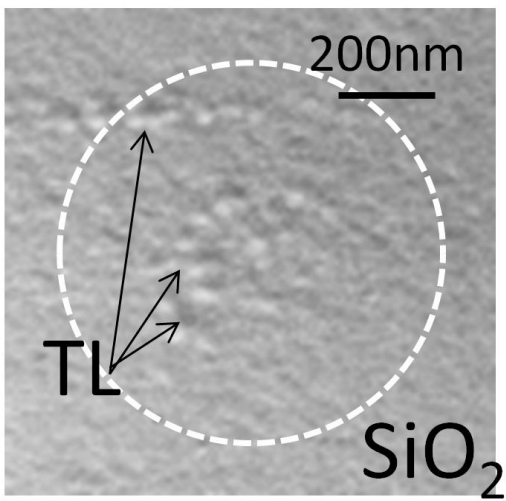
$R = 11.0\ \mu m$

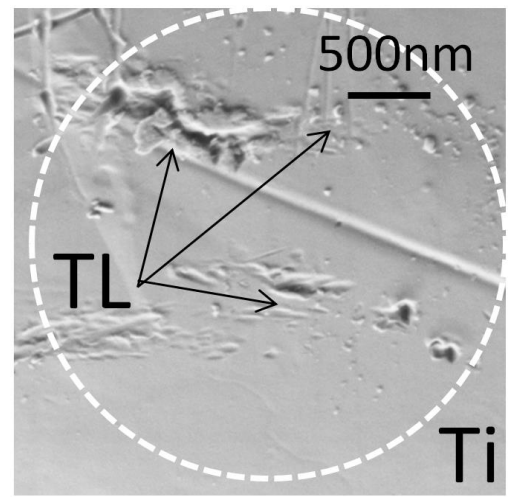
$R = 25.0\ \mu m$

**Figure S3. SEM micrographs of the contact area of each bead**. They confirm the presence of the tribotransferred flakes. Micrographs agree with AFM topographies of Figure 3 (main text).

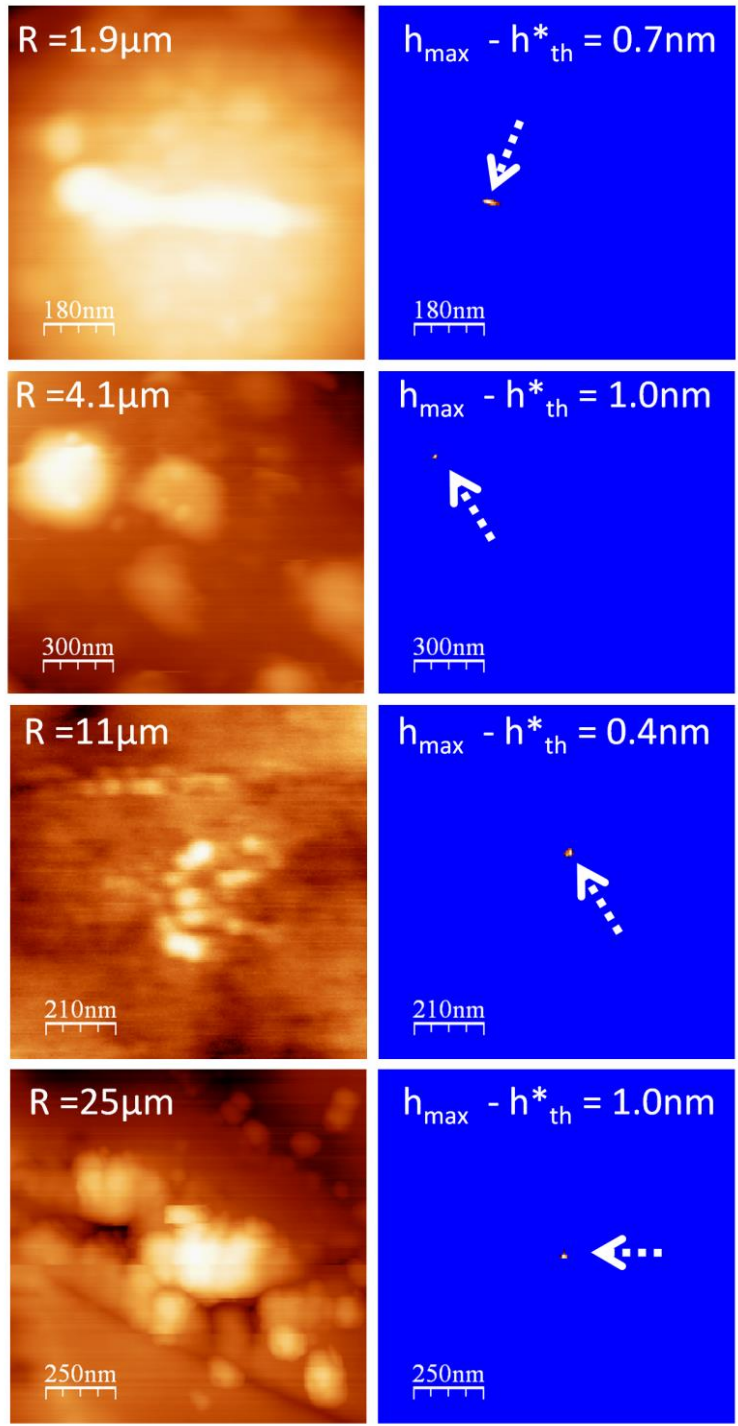

**Figure S4. Estimation of the area of the topographically-highest nanoasperity for each colloidal probe.** The area of the topographically-highest nanoasperity was roughly estimated by processing the topographies from Figure 3 of the main text (Figure S4, left column) with the "flooding" option of the WSXM software. This method is based on setting all the image values below a given threshold value $h_{th}$ (selected by the user) to a constant value. The resulting image resembles a picture of a flooded region of land (Figure S4, right column). Once the process is finished WSXM can calculate the area $A_{isl}(h_{th})$ of the "drought islands". The threshold value $h_{th}$ was varied for each topography in small steps of $0.1 - 0.2$nm below the maximum surface height $h_{max}$, to find out the threshold $h_{th}^*$ separating the regime of single-asperity contact ($0 < h_{th} \leq h_{th}^*$, only one island in the "flooded topography") from a multi-asperity contact ($h_{th} > h_{th}^*$, two or more islands in the "flooded" topography ). The area of the topographically-highest nanoasperity was estimated as $A_{isl}(h_{max} - h_{th}^*)$. We conventionally assumed that $h_{th}^*$ values that are more than 1nm apart from $h_{max}$ (i.e. $h_{max} - h_{th}^* > 1$nm) are unphysical. In fact the penetration depth of a nanoasperity into graphite is expected to be of only $\delta \sim 0.3 - 0.6$nm at the normal load $F_N = 100$nN (from Hertzian contact theory $\delta = (F_N^2/E^2 R)^{1/3}$ with the effective Young modulus $E = 29.5$GPa and a tentative curvature radius $R = 50 - 300$nm for the nanoasperity). In such specific situations the area of the topographically-highest nanoasperity was assumed to be $A_{isl}(h_{max} - 1\text{nm})$. We obtained that $A_{isl}$ is $\sim 0.3 - 0.6 \times 10^3$nm$^2$ for all the beads. White arrows in the "flooded topographies" highlight the location of the single "drought island".

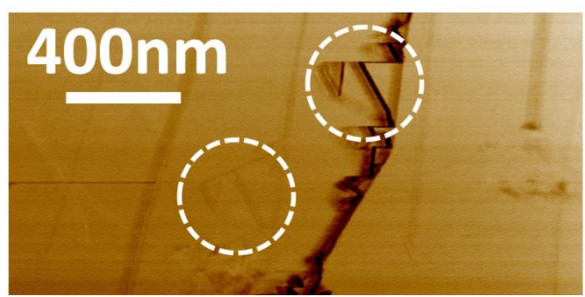
$R = 1.9$ μm

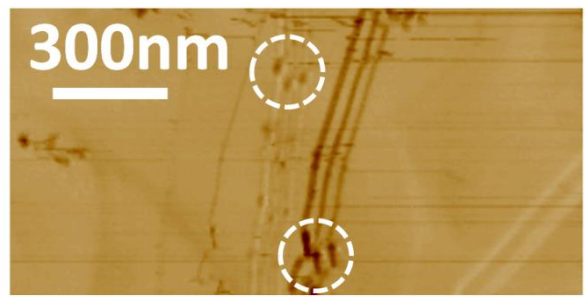
$R = 4.1$ μm

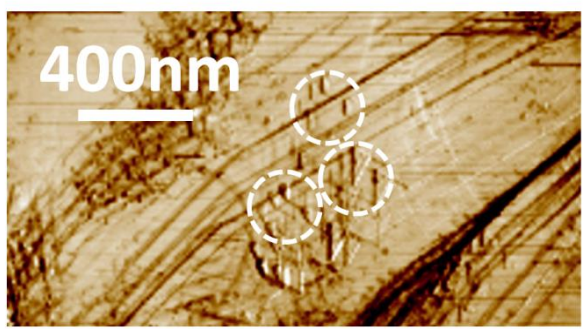
$R = 11.0$ μm

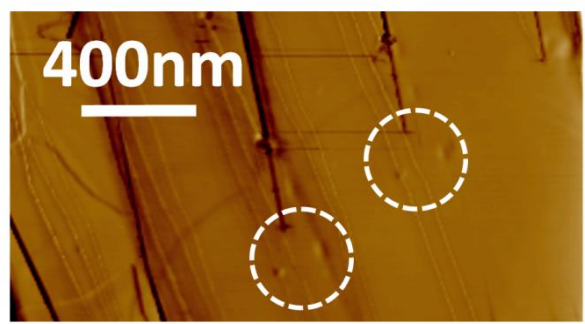
$R = 25.0$ μm

**Figure S5**. **Multiple-tip effects originated by a rough TL.** Examples of multiple-tip patterns in lateral force maps acquired with the TL/HOPG contact. Besides such patterns, multiple step edges occur.

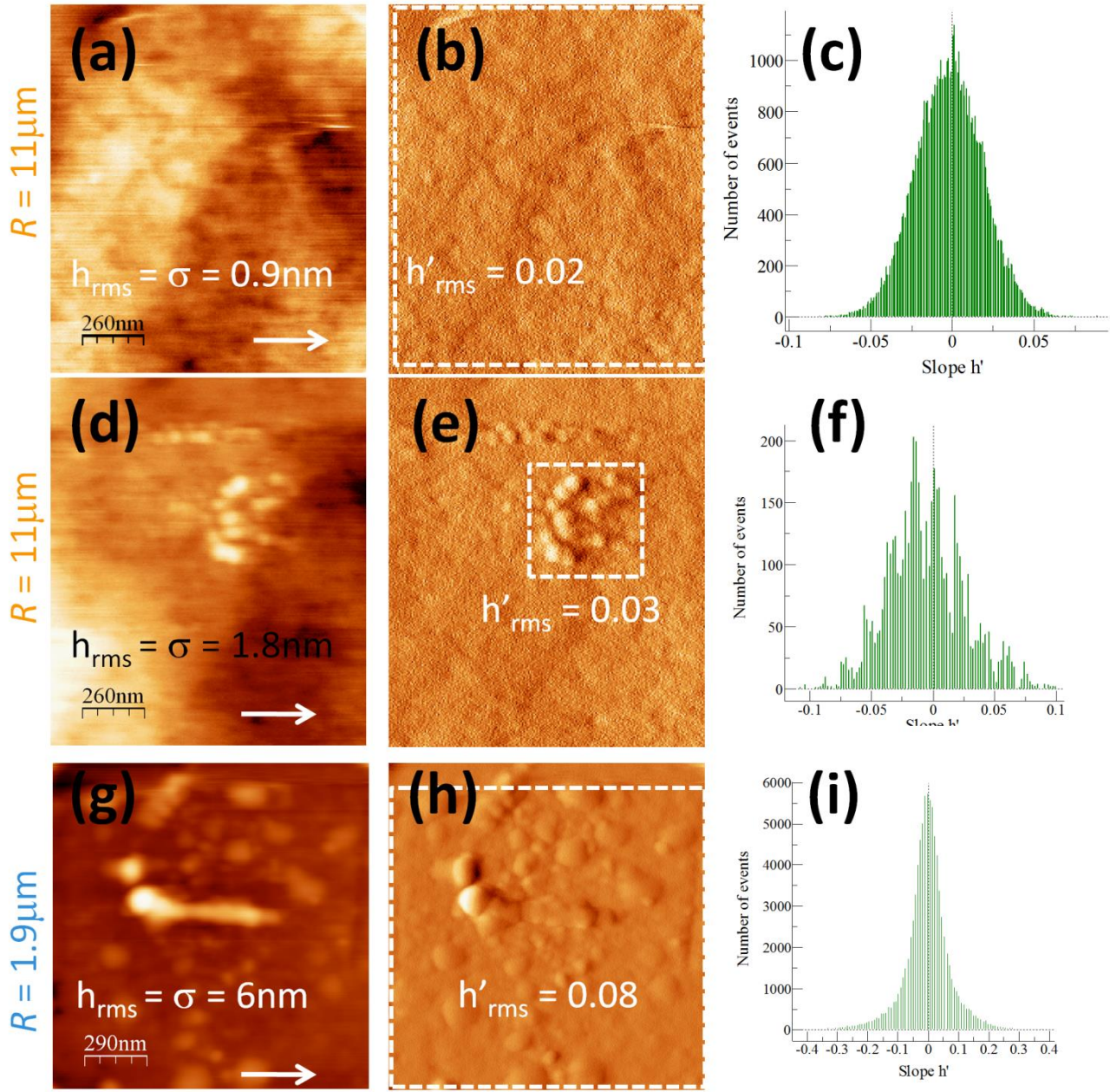

**Figure S6. Evaluation of the average local slope $h'_{rms}$ from AFM topographies.** (a-c). Pristine morphology of the bead with $R = 11\mu m$. From the topography map $h$ in (a), the map of local slopes $h'$ (along the white arrow direction) is obtained by finite differences (in (b)). Root mean square values are calculated for both maps. In (c) is the histogram of $h'$. (d-f) the same as above, but for the bead with $R = 11\mu m$ after the TL formation. (g-i) the same as in (d-f), for the bead with $R = 1.9\mu m$ and after TL formation. **Evaluation of the critical load $N_C$.** According to Pastewka et al. Appl. Phys. Lett. 2016, 108 (22), in the nonadhesive hard-wall repulsion limit ($1/\kappa = 0.5$), with $E^* = 30$GPa (HOPG-dominated interfacial deformation), and $R = 1.9\mu m$, $h'_{rms} = 0.08$ one gets $N_C \sim 120\mu N$. Also, for $R = 11\mu m$ and $h'_{rms} = 0.03$, one has $N_C \sim 210\mu N$. Hence $F_N/N_C \sim 5 - 8 \times 10^{-4}$ at $F_N = 100$nN and $F_N/N_C < 1 \times 10^{-4}$ for $F_N \sim 10$nN.

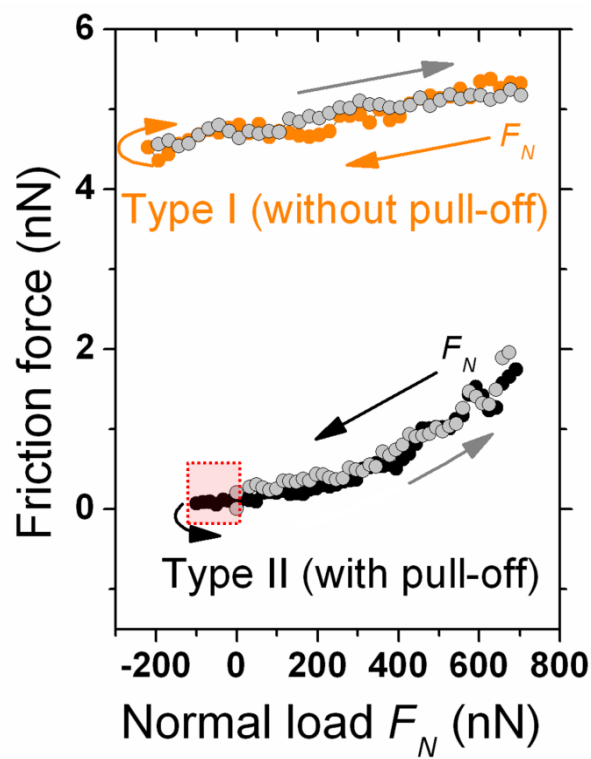

**Figure S7. Effect of contact loading/unloading on TL/HOPG friction characteristics.** Representative $F_f$ vs $F_N$ curves, of 'type I' and II, acquired with a $SiO_2$ colloidal bead via unloading-loading cycles (scan range $11 \times 11 nm^2$, $v = 33 nm/s$). Straight arrows highlight the $F_N$ variation along the unloading and loading branches respectively. Curved arrows indicate the turn point separating the two branches. Hysteresis is remarkably small for both characteristics. Adhesion hysteresis is discerned for the 'type II' curve (red square), as in this case the unloading branch extends up to the pull-off point i.e. the unloading-loading cycle implies contact rupture and formation. Noteworthy, the small hysteresis of both characteristics confirms that the response of the TL/HOPG contact junction is not affected by the specific procedure used to ramp $F_N$.

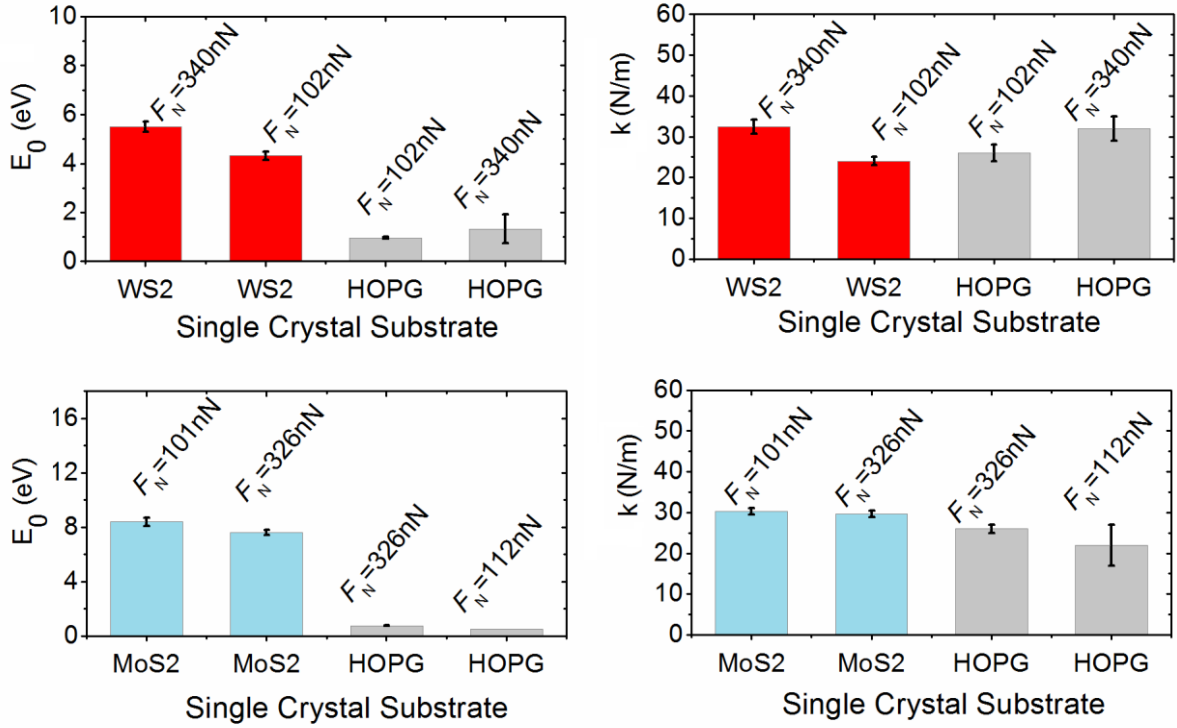

**Figure S8. Evaluation of potential corrugation $E_0$ and contact stiffness $k$ for sliding hetero-junctions, formed by contacting the tribo-induced graphitic TL with transition metal dichalcogenides.**

Top graphs: $E_0$ and $k_{eff}$ for the TL/WS$_2$ hetero-junction (red bars) are compared with the same quantities evaluated for the TL/HOPG homo-junction (grey bars; this is the superlubric starting condition, see main text). Two values of $F_N$ are taken into account (102nN and 340nN). One notes that $k(TL/WS_2) \sim k(TL/HOPG)$ whereas $E_0(TL/WS_2) \sim 5 \times E_0(TL/HOPG)$, that implies $\eta(TL/WS_2) \sim 3 \times \eta(TL/HOPG) \sim 6$.

Bottom graphs: $E_0$ and $k_{eff}$ for the TL/MoS$_2$ hetero-junction (light blue bars) are compared with the same quantities evaluated for the TL/HOPG homo-junction (grey bars; superlubric starting condition). In this case $k(TL/MoS_2) \sim k(TL/HOPG)$ whereas $E_0(TL/MoS_2) \sim 12 \times E_0(TL/HOPG)$, and $\eta(TL/MoS_2) \sim 4 \times \eta(TL/HOPG) \sim 8$.